\input{aipcheck}

\documentclass[
    ,final            
  ]
  {aipproc}

\layoutstyle{6x9}

\begin{document}

\title{Mapping the QCD Phase Transition with Accreting Compact Stars}


\author{David Blaschke}{
  address={Institute for Theoretical Physics, University of Wroclaw, 
Max-Born pl. 9, 50-204 Wroclaw, Poland}
,altaddress={Bogoliubov Laboratory for Theoretical Physics, JINR Dubna, 
Joliot-Curie str. 6, 141980 Dubna, Russia } 
}

\author{Gevorg Poghosyan}{
  address={Forschungszentrum Karlsruhe GmbH, Hermann-von-Helmholtz-Platz 1,
D-76344 Eggenstein-Leopoldshafen, Germany}
}

\author{Hovik Grigorian}{
  address={Laboratory for Information Technologies, JINR Dubna, 
Joliot-Curie str. 6, 141980 Dubna, Russia}
}

\begin{abstract}
We discuss an idea for how accreting millisecond pulsars could contribute 
to the understanding of the QCD phase transition in the  
high-density nuclear matter equation of state (EoS).
It is based on two ingredients, the f{i}rst one being a ``phase diagram''
of rapidly rotating compact star conf{i}gurations in the plane of 
spin frequency and mass, determined with state-of-the-art 
hybrid equations of state, allowing for a transition to color superconducting 
quark matter. The second is the study of spin-up and accretion evolution in 
this phase diagram.
We show that the quark matter phase transition leads to a 
characteristic line in the $\Omega-M$ plane, the phase border between 
neutron stars and hybrid stars with a quark matter core.
Along this line a change in the pulsar's moment of inertia entails a waiting 
point phenomenon in the accreting millisecond X-ray pulsar (AMXP) evolution: 
most of these objects should therefore be found along the phase border in the  
$\Omega-M$ plane, which may be viewed as the AMXP analog of the main 
sequence in the Hertzsprung-Russell diagram for normal stars.
In order to prove the existence of a high-density phase transition in the 
cores of compact stars we need population statistics for AMXP's with
suff{i}ciently accurate determination of their masses and spin frequencies.  
\end{abstract}

\maketitle


\section{Introduction}

Accreting compact stars in low-mass binary systems undergo a 
stage with disc accretion leading to both spin-up and mass increase.
Initial indications for spin frequency clustering in low-mass X-ray
binary (LMXB) systems, reported by measurements with the Rossi-XTE, have lead 
to the interpretation of such a correlation as 
a waiting-point phenomenon 
where stellar conf{i}gurations cross the border between pure neutron stars
and hybrid stars in the spin frequency-mass plane 
\cite{Glendenning:2000zz}.
A systematic analysis of the critical line for a deconf{i}nement phase 
transition in the ``phase diagram'' for accreting compact stars 
\cite{Poghosyan:2000mr} has revealed that the suggested population clustering
due to the phase transition will instead lead to mass clustering.
For homogeneous interiors like in the case of strange stars, however, such an 
effect shall be absent \cite{Blaschke:2001th}. 
For generic polytropic forms of the EoS of quark and 
hadronic matter, the relationship between  softness or hardness of the EoS and
the structure of this phase diagram has been demonstrated in Ref. 
\cite{Grigorian:2002ih}. 

In the present contribution we give an updated view on these ideas, based on a 
recently developed hybrid EoS \cite{Klahn:2006iw}
which fulf{i}lls constraints from compact star observations and heavy-ion 
collision (HIC) experiments \cite{Klahn:2006ir}.
This example demonstrates the compatibility of an onset of deconf{i}nement in 
stars with typical masses $\sim~1.4~M_\odot$ 
with a high maximum mass $\sim~2.0~M_\odot $. 
We speculate that the class of objects for which mass clustering due to a 
phase transition in their interior applies might include also most of the 
first-born neutron stars in low-kick, non-excentric double neutron star as well
as in pulsar-white dwarf systems which have undergone a mass accretion stage
\cite{Heuvel:2007zp}.

Upon further elaboration and proper selection of the pulsar population, 
the mass clustering phenomenon could prove to be a direct observation 
of a phase transition in compressed nuclear matter, such as the QCD 
chiral symmetry restoration transition.

\section{Hybrid EoS: Mass-Radius vs. Flow Constraints}

One of the most challenging tasks in fundamental nuclear and particle physics 
is the delineation of the border between hadronic matter and quark-gluon matter
in the temperature-density plane: the QCD phase diagram.
At high temperatures and low baryon densities, numerical simulations of QCD
as a lattice gauge theory indicate that both, chiral symmetry restoration and 
deconf{i}nement are crossover transitions and their critical temperatures 
coincide at a value of $T_\chi=T_d=196$ MeV, obtained by the 
Bielefeld-Brookhaven-Columbia-Riken collaboration 
\cite{Cheng:2007jq}\footnote{We want to remark that a 
considerably lower value of $T_\chi=151$ MeV has been reported by Fodor et al. 
\cite{Fodor:2007ue}. The discrepancy to Cheng et al.\cite{Cheng:2007jq} is not
yet resolved.}. 
This value is consistent with a statistical model analysis of
the freeze-out temperature $T_f = 160$ MeV from hadron production in nuclear 
collisions at RHIC for the highest presently available c.m.s. energy of
$\sqrt{s}=200$ GeV \cite{Andronic:2005yp,Cleymans:2006qe}.

At f{i}nite densities and $T\approx 0$, the situation is quite different. 
Both lattice QCD simulations and heavy-ion collision experiments cannot access
this region and many questions are yet unanswered, like:
\begin{itemize}
\item[(i)] Is (are) the transition(s) of f{i}rst order so that there must be a 
critical endpoint in the QCD phase diagram \cite{Stephanov:1999zu}? 
\item[(ii)] Is quark matter at low $T$ a color superconductor and how
does such a property manifest itself \cite{Alford:1997ps}? 
\end{itemize}
Other points of discussion which we will not discuss in detail here are:
\begin{itemize}
\item[(iii)] Do all quark flavors appear simultaneously or rather sequentially 
\cite{Ruester:2005jc,Blaschke:2005uj,Abuki:2005ms,Blaschke:2008br}?
\item[(iv)] Do chiral and deconf{i}nement transition happen at the same 
critical chemical potential $\mu_{\chi}=\mu_d$ or is there a so-called 
{\em quarkyonic phase} \cite{McLerran:2007qj}?
\end{itemize}
At present, the description of cold, dense matter phases within QCD is out
of reach. 
Therefore, effective models for quark matter like those of the 
Nambu--Jona-Lasinio (NJL) type \cite{Buballa:2003qv} may help to obtain 
quantitative estimates for the dense matter EoS and phase transitions, 
once the free parameters can be f{i}xed. 
In this situation, astrophysical observations of 
compact stars 
may provide constraints for the behavior of matter under high compression and 
isospin asymmetry, complementary to data from lattice QCD and heavy-ion 
collisions.   

\begin{figure}[h!] 
\includegraphics[height=0.49\textwidth, angle=-90]{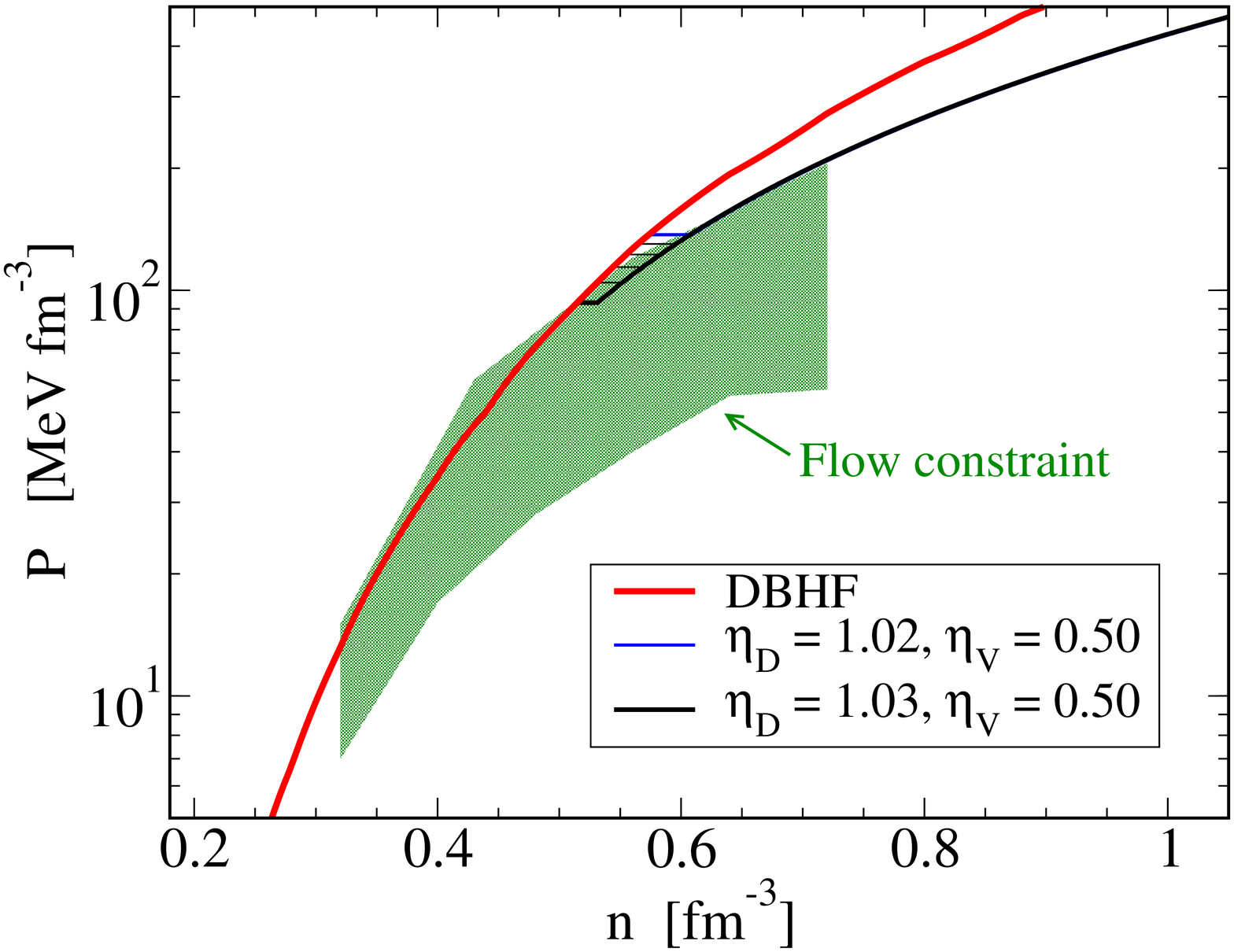} 
\includegraphics[height=0.49\textwidth, angle=-90]{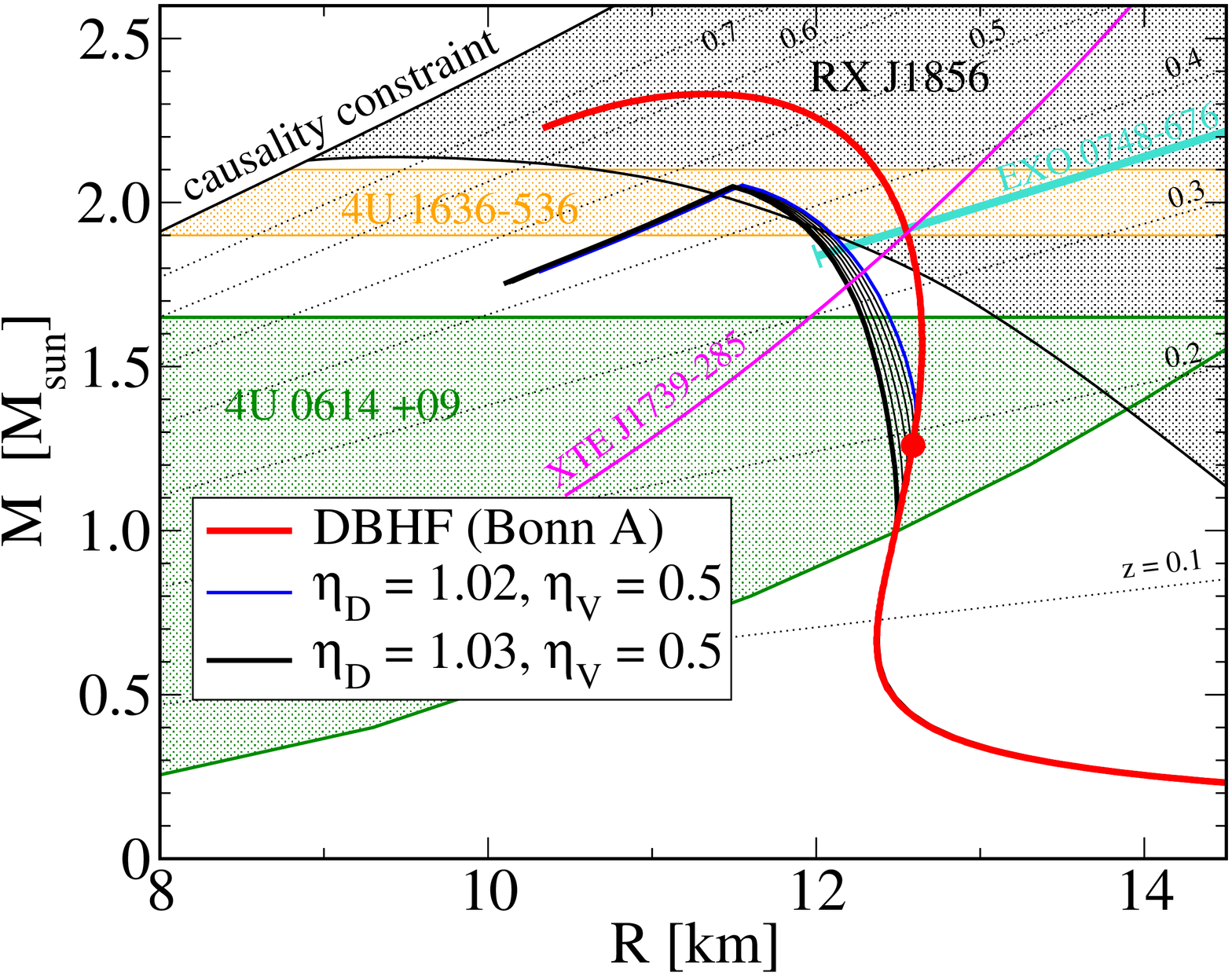} 
\caption{ 
{\em Left:} Hybrid EoS with DBHF nuclear matter and color superconducting 
NJL quark matter f{i}tted to obey the f{l}ow constraint 
\cite{Danielewicz:2002pu}.
{\em Right:}  Mass-Radius constraints from thermal radiation of the 
isolated neutron star  RX J1856.5-3754 (grey hatched region) and from 
QPOs in the LMXB's 4U 0614+09  (green hatched area, the ``wedge'') and 
4U 1636-536 (orange hatched region) 
which shall be  regarded as separate conditions to the EoS, see 
Ref.~\cite{Klahn:2006iw}. 
The controversial interpretation that the 1122 Hz pulsations for XTE J1739-285 
correspond to its spin frequency \cite{Kaaret:2006gr}
entails a stringent mass-radius constraint for this compact star
\cite{Lattimer:2006xb}.
The theoretical $M-R$ relations correspond to hybrid star conf{i}gurations 
with a DBHF hadronic shell and a color superconducting  NJL
quark matter core. The underlying hybrid EoS is the isospin-asymmetric 
generalization of that in the left panel, under $\beta$-equilibrium with 
electrons and muons.
Shown are also lines of constant surface redshift $z$.   
\label{f{i}g:M-R}  } 
\end{figure}  

In the spirit of two-phase models for the dense matter EoS, the nuclear
matter EoS can be described by the ab-initio Dirac-Brueckner-Hartree-Fock
(DBHF) approach \cite{Fuchs:2003zn,vanDalen:2004pn} using the Bonn-A 
nucleon-nucleon potential and the quark matter EoS is given by a 
color superconducting three-flavor NJL model \cite{Blaschke:2005uj} 
augmented by a vector meanf{i}eld contribution \cite{Klahn:2006iw} which 
stiffens the deconf{i}ned phase. 
In the left panel of F{i}g.~\ref{f{i}g:M-R} we illustrate how the coupling 
strengths in the scalar diquark ($\eta_D$) and in the vector meson ($\eta_V$) 
channels as free parameters of the NJL model description can be f{i}xed by the
requirement that the isospin-symmetric hybrid EoS resulting from a Maxwell 
construction with the DBHF EoS is as stiff as possible but still compatible
with the constraint region from HIC flow data at AGS and SIS energies
\cite{Danielewicz:2002pu}.
Resulting preferable parameter values are $\eta_V=0.5$ and $\eta_D$ in the 
range $\eta_D=1.02 ... 1.03$.
The onset of the hadron-to-quark matter transition in symmetric nuclear matter
is at a density of $n_d=0.55$ fm$^{-3}=3.5~n_0$, where $n_0=0.16$ fm$^{-3}$ 
denotes the nuclear matter saturation density.
The generalization of this hybrid EoS to the isospin-asymmetric case under
$\beta$-equilibrium with electrons and muons can be used to {\em predict}
the critical mass of a neutron star for the onset of deconf{i}nement in its
interior in the range of $M_{\rm crit} \sim 1.35 \dots 1.0 ~M_\odot$, see also
\cite{Klahn:2006iw}.
The corresponding critical density for the hadron-to-quark matter transition 
in compact stars is at about $n_{\rm crit}=0.4$ fm$^{-3}=2.5~n_0$. 
In the right panel of F{i}g.~\ref{f{i}g:M-R} we show the sequences of hadronic
compact stars (red line) and the hybrid star sequences with color 
superconducting quark matter cores (black and blue lines).
We would like to stress two results: (i) we predict that compact stars in the 
typical mass range of $1.35\pm 0.1~M_\odot$ may have quark matter interiors, 
and (ii) the maximum mass for compact stars with quark matter interiors 
exceeds $2~M_\odot$. 
The latter result disproves the claim \cite{Ozel:2006bv} that the 
occurrence of deconfined quark matter in compact stars would necessarily 
lead to a softening of the EoS which excludes the possibility of hybrid star 
configurations with large masses and radii, see also Ref.~\cite{Alford:2006vz}.

However, this demonstrates only that the possibility of compact stars having 
quark matter interiors cannot be excluded by the measurement of a large mass. 
What does it take to discover a high-density phase transition by astrophysical
observations? 
With the newly developed hybrid star EoS at hand we want to revisit the 
suggestion that a clustering of frequencies 
\cite{Glendenning:2000zz} and/or masses \cite{Poghosyan:2000mr}
in the population of compact stars in LMXB's may be seen as a signal for 
deconfinement and add a new twist to this hypothesis by suggesting to include 
double neutron stars into these considerations!

\section{Population clustering as a signal for deconfinement}

Our investigation is based on the classif{i}cation of rotating compact
star conf{i}gurations in the plane of angular velocity $\Omega$ and
baryon number $N$ (or gravitational mass $M$), the so called 
{\it phase diagram} for compact stars \citep{Blaschke:2000xy}.
It is defined by at least three lines: the maximum frequency 
$\Omega_{\rm max}(N)$ for which stable rotation without mass shedding
can be sustained, the maximum baryon number $N_{\rm max}(\Omega)$ the star
can carry without undergoing gravitational collapse and 
the critical line $N_{\rm crit}(\Omega)$, which
separates the region of quark core configurations from the hadronic ones,
see Fig.~\ref{f{i}g:mass-clus-high}.
It has been shown that the latter line is correlated with
the local maxima of the moment of inertia with respect to changes of
the baryon number at given $\Omega$ due to the change of the internal
structure of the compact object at the deconf{i}nement phase transition.
Therefore, we expect that the rotational behavior of these objects
changes in a characteristic way when this line is crossed. 
Early suggestions of a deconfinement signal following from this 
characteristic behavior have considered the spin-down of isolated radio
pulsars without mass accretion for which a characteristic deviation of the 
braking index from the value $n=3$ for dipole emission shall signal the 
transition \cite{Glendenning:1997fy,Chubarian:1999yn}.
It has also been noted that the star has to spend about $10^8$
yr for crossing the configuration border in the phase diagram, the typical
time it takes the star to loose by dipole emission the amount of angular 
momentum $\Delta J= \Omega \Delta I $ which corresponds to the change in 
the moment of inertia $ \Delta I$ sue to the phase transition in the star's 
interior.
The consequence will be an increase of the population of stars at this
critical line which could be observed. Provided that a suff{i}ciently
large number of accretors will be discovered and their masses and spin
frequencies \citep{Poghosyan:2000mr} could be determined.
Then the phase transition would reveal itself by a population clustering 
along the line $N_{\rm crit}(\Omega)$ in the phase diagram, which according 
to the above results for nonrotating stars should be well separated  
from the black hole limit $N_{\rm max}(\Omega)$. 

The main problem with this signal is the shape of the critical line in the 
phase diagram which for typical hybrid EoS \cite{Grigorian:2002ih} would 
suggest a mass clustering \cite{Poghosyan:2000mr} rather than a 
frequency clustering \cite{Glendenning:2000zz}.
In order to measure the star mass, however, one would need a companion star 
which for isolated pulsars is absent! 

Therefore, the suitable population for which this statistical 
phase transition test shall be applicable are compact stars in binary systems 
with mass transfer via Roche-lobe overflow.
There are two such systems which we want to focus on:
(i) LMXB's \cite{vanderKlis:2000ca} and (ii) double neutron stars 
(DNS's), see \cite{Heuvel:2007zp} for a recent discussion.


The spin evolution of a compact star under mass accretion from a low-mass 
companion star can be regarded as a sequence of stationary states of
conf{i}gurations (points) in the phase diagram. 
It is governed by the change in angular momentum of the star

\begin{equation} 
\label{djdt} 
\frac{d}{dt} (I(N,\Omega)~ \Omega)= K_{\rm ext}~,
~~K_{\rm ext}= \sqrt{G M \dot M^2 r_0}- N_{\rm out}~,
\end{equation} 
where $K_{\rm ext}$ denotes the external torque due to both the specif{i}c 
angular momentum transfered by the accreting plasma and the magnetic plus
viscous stress given by $N_{\rm out}=\kappa \mu^2 r_c^{-3}$, 
$\kappa=1/3$ \citep{lipunov}. For a star with radius
$R$ and magnetic f{i}eld strength $B$, the magnetic moment is given by
$\mu=R^3~B$ and
$r_c=\left(GM/\Omega^2\right)^{1/3}$ is the co-rotating radius, see
\cite{Glendenning:2000zz,Poghosyan:2000mr} and references therein for details.
From Eq.  (\ref{djdt}) follows the evolution equation for the angular velocity

\begin{equation} 
\label{odoto} 
\frac{d \Omega}{d t}= 
\frac{K_{\rm ext}(N,\Omega)- K_{\rm int}(N,\Omega)} 
{I(N,\Omega) + {\Omega}({\partial I(N,\Omega)}/{\partial \Omega})_{N}}~,~~ 
K_{\rm int}(N,\Omega)=\Omega\dot N 
\left(\frac{\partial I(N,\Omega)}{\partial N}\right)_{\Omega}~. 
\end{equation}
 
Solutions of (\ref{odoto}) are trajectories in the $\Omega - N$ plane
describing the spin evolution of accreting compact stars.  
Since $I(N,\Omega)$ exhibits characteristic
functional dependences \citep{Blaschke:2000xy} at the deconf{i}nement phase
transition line $N_{\rm crit}(\Omega)$ we expect observable
consequences in the $\dot P - P$ plane when this line is crossed.

In our model calculations we assume that both the mass accretion and
the angular momentum transfer processes are slow enough to justify the
assumption of quasistationary rigid rotation without convection.  The
moment of inertia of the rotating star can be def{i}ned as $I(N,\Omega)=
J(N,\Omega)/\Omega~$, where $J(N,\Omega)$ is the angular momentum of
the star.  For a more detailed description of the method and analytic
results we refer to \citep{Chubarian:1999yn} and references therein.
The time dependence of the baryon number for the constant accreting
rate $\dot N$ is given by 
$N(t)=N(t_0)+ (t-t_0)\dot N$
and for the magnetic f{i}eld of the accretors we consider the exponential 
decay 
$B(t)=[B(0) - B_{\infty}]\exp(-t/\tau_B)+ B_{\infty}$.
We solve the equation for the spin-up evolution (\ref{odoto}) of the
accreting star for decay times $\tau_B = 10^9$ yr and
initial magnetic f{i}elds in the range $0.2 \leq B(0){\rm [TG]}\leq 4.0
$.  The remnant magnetic f{i}eld is chosen to be
$B_\infty=10^{-4}$TG\citep{page}, where 1 TG= $10^{12}$ G.

The question arises whether there is a characteristic feature in the
spin evolution when the trajectories traverse the
critical phase transition line.
In order to perform
a more quantitative discussion of possible signals of the deconf{i}nement
phase transition we investigate the spin-up evolution for stars with
$N(0)=1.4~N_\odot$ and $\Omega(0)=1$ Hz in the initial state. In the case of
high accretion rate ($\dot N=10^{-8}N_\odot$/yr, e.g.  for Z sources)
and long-lived magnetic f{i}eld ($\tau_B=10^9$yr) there is a dip in the period
derivative when the star evolves into the quark core region of the phase 
diagram. This feature can be quantif{i}ed
by the distribution of a {\it waiting time} 
$\tau=\left|P/\dot P\right|=\Omega/ \dot\Omega$
in the $\Omega - N$ plane.  
\begin{figure}[htb]     
\includegraphics[width=0.9\textwidth,height=0.5\textwidth,angle=0]{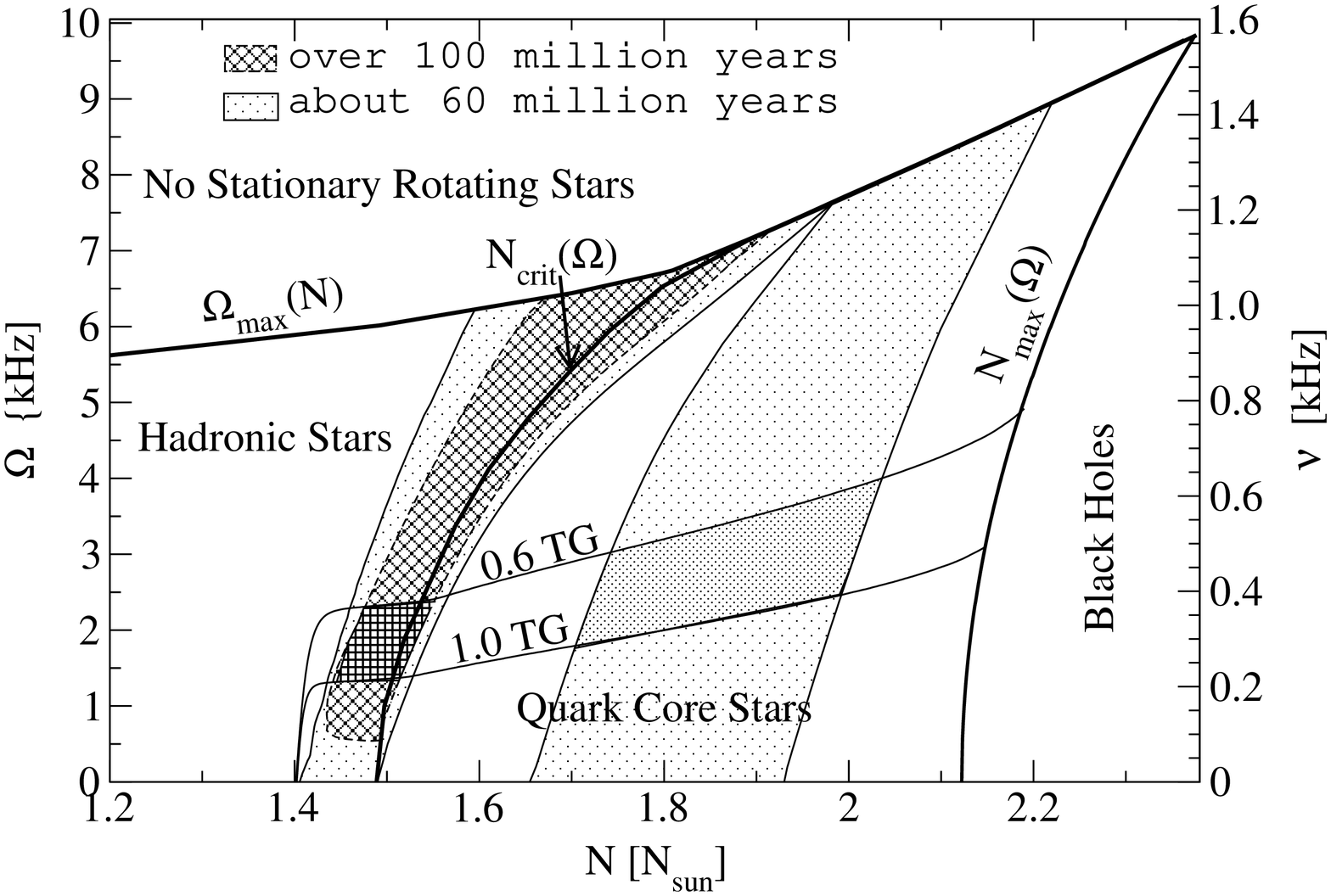}
\caption{
Regions of {waiting times} in the phase
diagram for compact hybrid stars for a magnetic field decay time of 
$\tau_B=10^9$ yr and a constant mass accretion rate of 
$\dot N=10^{-8}~N_\odot$/yr.  
For an estimate of a population statistics we show the region of evolutionary
tracks when the interval of initial magnetic field values is restricted
to $0.6\leq B(0)[{\rm TG}] \leq 1.0$.  Note that the probability of
finding a compact star in the phase diagram is enhanced in the vicinity
of the critical line for the deconfinement phase transition $N_{\rm
crit}(\Omega)$ by at least a factor of two relative to all other
regions in the phase diagram.
The correlation of the distribution of objects with the line for a 
deconf{i}nement  phase transition could be interpreted as a waiting point 
phenomenon during the accretion evolution of the compact stars 
(mass clustering), see \cite{Poghosyan:2000mr}.
\label{f{i}g:mass-clus-high} }      
\end{figure}     

In F{i}g.  \ref{f{i}g:mass-clus-high} we show contours of waiting time 
regions in the phase diagram.  
The region of longest waiting times is located in a narrow branch
around the phase transition border and does not depend on the evolution
scenario after the passage of the border, when the depopulation occurs
and the probability to f{i}nd an accreting compact star is reduced.
Another smaller increase of the waiting time and thus a population
clustering could occur in a region where the accretor is already a 
quark core star.
For an estimate of the magnetic field influence we show 
in F{i}g.  \ref{f{i}g:mass-clus-high} also the region of
evolutionary tracks when the values of initial magnetic f{i}eld vary
within $0.6\leq B(0)[{\rm TG}] \leq 1.0$.

As a strategy of search for QCSs we suggest to select from the LMXBs
exhibiting the QPO phenomenon those accreting close to the Eddington
limit \citep{heuvel} and to determine simultaneously the spin frequency
and the mass \citep{Lamb:2000wd} for suff{i}ciently many of these objects.  
The emerging statistics of accreting compact stars should then exhibit the
population clustering shown in F{i}g. \ref{f{i}g:mass-clus-high}  when a
deconf{i}nement transition is possible.  If a structureless distribution
of objects in the $\Omega - N$ plane will be observed, then no f{i}rm
conclusion about quark core formation in compact stars can be made as, e.g.,
for strange stars \cite{Blaschke:2001th}. 
There is a problem with the simultaneous analysis of masses and spin 
frequencies for LMXB's which is still rather dependent on the model 
employed \cite{stella,MLP}.

\begin{figure}[htb]     
\includegraphics[width=0.5\textwidth,height=0.9\textwidth,angle=-90]{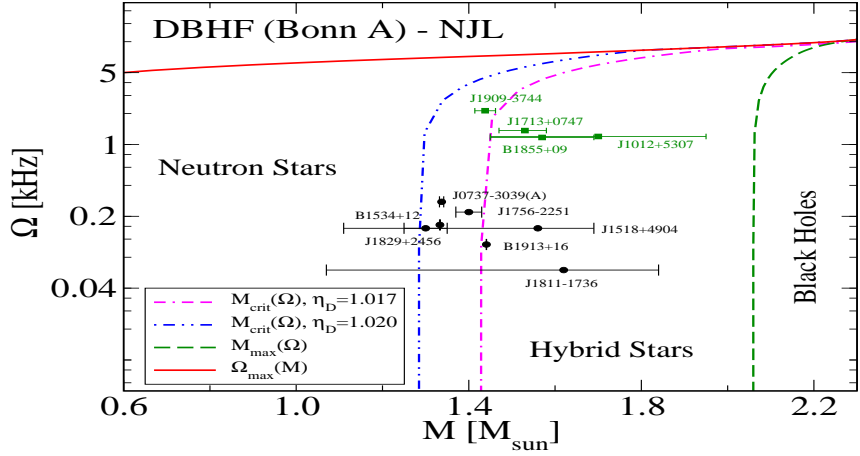} 
\vspace{-1cm}
\caption{
``Phase diagram'' for compact stars in the plane of rotation frequency 
$\Omega$ vs. mass $M$. Stable stars are found in the region bordered by 
the maximum rotation frequency $\Omega_{\rm max}(M)$ for which mass-shedding 
from the star equator occurs and the maximum mass $M_{\rm max}(\Omega)$ for
which the gravitational instability against collapse to a black hole sets in.
The dash-dotted (dash-double-dotted) line marks the onset of a quark
matter core in the stars interior for a diquark coupling parameter 
$\eta_D=1.017$ ($\eta_D=1.02$), see \ref{f{i}g:M-R} 
for the corresponding
mass-radius relations of the nonrotating star sequences. 
The data with error bars correspond to binary radio pulsars (black dots) 
\cite{Heuvel:2007zp} and
neutron stars with a white dwarf binary (green squares) \cite{Lattimer:2006xb}.
The correlation of the distribution of objects with the line for a 
deconf{i}nement  phase transition could be interpreted as a waiting point 
phenomenon during the accretion evolution of the compact stars 
(mass clustering), see \cite{Poghosyan:2000mr}.
\label{f{i}g:mass-clus-low} }      
\end{figure}     

Following van den Heuvel \cite{Heuvel:2007zp}, we may suggest to apply our 
approach also to the population of the first-born stars in double neutron 
star systems with low excentricity and pulsar-white dwarf double systems,
as they should have undergone a mass accretion stage similar to the one
described above which is a prerequisite for the applicability of the 
population clustering signal of deconfinement.  
For the specif{i}c example of the hybrid EoS discussed above with a DBHF 
hadronic phase and a color superconducting stiff quark matter phase, we show 
in Fig.~\ref{f{i}g:mass-clus-low} the phase diagram of rotating compact stars 
together with the masses and spin frequencies obtained for double neutron star 
systems.  
We observe an interesting correlation of the distribution of these objects 
with the critical deconf{i}nement phase transition line for diquark coupling 
in the range $\eta_D=1.017 ... 1.02$. 
There might be other reasons for the mass clustering of spinning pulsars 
\cite{Heuvel:2007zp}, but the suggestion to relate it to a phase transition 
in the compact star interior can not be excluded. 

\section{Summary}
The model independent result of our study
is that a population clustering in the phase diagram for accreting
compact stars shall measure the critical line $N_{\rm crit}(\Omega)$
which separates neutron stars from hybrid stars where the shape of this curve
can discriminate between different models of the nuclear EoS at high
densities.

For the new hybrid equation of state discussed in this contribution, 
we expect the suggested population clustering 
as a signal of the deconf{i}nement transition in the mass range 
$M\sim 1.3 \dots 1.4~M_\odot$ which might therefore serve as one aspect for
the explanation of the well-known mass clustering in double neutron stars
in that same mass range.
The above rather fresh ideas suggest that future observational programs may
contribute to unraveling most actual problems of fundamental physics of dense 
baryonic matter.


\begin{theacknowledgments}

Thanks go to Thomas Kl\"ahn and Fredrik Sandin for providing the hybrid
EoS; to Christian Fuchs and Stefan Typel for the hadronic part of the EoS.
D.B. acknowledges discussions with Jim Lattimer and Cole Miller on this 
paper and the ESF Research Networking Programme ``CompStar'' for 
supporting his participation at this conference. He is also grateful for the
hospitality of the Institute for NUclear Theory at the University of Washington
and for partial support from the Department of Energy during the completion of 
this work.

\end{theacknowledgments}

\end{document}